\documentclass[aps,twocolumn]{revtex4}
\usepackage{graphicx}
\begin{document}
\newcommand{\D}{\displaystyle}
\title{Effect of anisotropic impurity scattering on a density of
states of a d-wave superconductor} 
\author{A. Maci\c{a}g, P. Pisarski, G. Hara\'n}
\affiliation{Institute of Physics, Politechnika
Wroc{\l}awska, Wybrze\.ze Wyspia\'nskiego 27, 50-370 Wroc{\l}aw,
Poland}

\begin{abstract}
We discuss the effect of an anisotropic impurity potential on the critical 
temperature, local density of states in the vicinity of a single impurity, 
and the quasiparticle density of states for a finite impurity concentration 
in a d-wave superconductor. Different scattering regimes are concerned.  
\end{abstract}
\maketitle

\section{Introduction}
Impurities make a useful tool in probing the ground state of
high-temperature superconductors.  
The most direct impurity probe of the superconducting state 
is provided by the scanning tunneling microscopy (STM) measurement 
of the local quasiparticle density of states (LDOS) around a single impurity. 
The STM images of ${\rm Bi_2Sr_2CaCu_2O_{8+\delta}}$ reveal a tetragonal 
symmetry representative for the d-wave state \cite{4,5,6,7}. In addition 
to a possible identification of the superconducting state, this 
experiment may shed light on the nature of the quasiparticle scattering 
process - issue, which is important to the interpretation of the 
impurity pair-breaking effect in the d-wave superconductor. 
It has been shown that the weak impurity-induced suppression of the critical 
temperature, which is unexpected for d-wave superconductivity \cite{8,9,10,11} 
can be understood assuming a momentum-dependent (anisotropic) impurity potential 
\cite{12,13,14,15}. In the present paper we discuss the effect of anisotropic 
impurity scattering on the critical temperature, local density of states in the 
vicinity of a single impurity, and the quasiparticle density of states for a 
finite impurity concentration in a d-wave superconductor. We establish how the 
internal structure of a defect, which reproduces $T_c$ suppression in the cuprates, 
alters the above quantities. We assume that the impurity potential factorizes 

\begin{equation}
\label{e1}
v\left({\bf k},{\bf k}'\right)=v_i+
v_af\left({\bf k}\right)f\left({\bf k}'\right)
\end{equation}

\noindent
where $v_i$, $v_a$ are isotropic and anisotropic scattering amplitudes,
respectively, and $f\left({\bf k}\right)$ is the anisotropy function that
vanishes after integration over the Fermi surface. 
We study the effect of anisotropy by referring to an isotropic impurity potential
$v_0$ that determines the scattering strength in both isotropic and anisotropic
scattering channels: $v_i=\alpha v_0$, and $v_a=\left(1-\alpha\right)v_0$; where
$0\le\alpha\le 1$ is a partition parameter. Scattering is isotropic for
$\alpha=1$ and purely anisotropic for $\alpha=0$. For the analysis of
different scattering limits we introduce a convenient parameter
$c=1/\left(\pi N_0 v_0\right)$, where $N_0$ is the density of
states per spin at the Fermi level in the normal state. 
Born scattering is then given by $c\gg 1$, and the unitary scattering limit 
is determined by $c=0$. In the calculations  
we assume a parabolic energy band and two-dimensional
superconductivity. We discuss the impurity effect
on the d-wave superconducting state determined by the order parameter
$\Delta\left({\bf k}\right)=\Delta_0 e\left({\bf k}\right)$,
where $e\left({\bf k}\right)=\sqrt 2\left(k^2_x-k^2_y\right)=\sqrt 2 cos2\phi$, 
and the amplitude of $\Delta\left({\bf k}\right)$ is defined as
$\Delta=\sqrt 2\Delta_0$.

\section{Critical temperature and order parameter suppression} 
The role of the anisotropy of the impurity potential is particularly clear 
for the potential (\ref{e1}) with an oscillating anisotropy function,  
which is given on the Fermi surface as $f\left({\bf k}\right)=\pm 1$. 
The critical temperature in the Born limit reads \cite{12,15} 

\begin{equation}
\label{e2}
\begin{array}{l}
\D\ln\frac{T_{c}}{T_{c_{0}}}=\left(\left<e\right>^{2}+
\left<ef\right>^{2}-1\right)
\left[\psi\left(\frac{1}{2}+\frac{\pi N_0\left(v_i^2+v_a^2\right)}
{2\pi T_{c}}\right)\right.\\ 
\\
\D\left. -\psi\left(\frac{1}{2}\right)\right]+
\left<ef\right>^{2}\left[\psi\left(\frac{1}{2}\right)\right.\\
\\
\D\left. -\psi\left(
\frac{1}{2}+\frac{\pi N_0\left(v_i^2+v_a^2\right)}{2\pi T_{c}}
\left(1-\frac{2v_iv_a}{v_i^2+v_a^2}\right)\right)\right]\\
\end{array}
\end{equation}

\noindent
and for unitary scattering is given by 

\begin{equation}
\label{e3}
\D\ln\frac{T_{c}}{T_{c_{0}}}=\left(\left<e\right>^{2}+
\left<ef\right>^{2}-1\right)
\left[\psi\left(\frac{1}{2}+\frac{2\Gamma}{2\pi T_{c}}\right)  
-\psi\left(\frac{1}{2}\right)\right]
\end{equation}

\noindent
where $T_{c_{0}}$ is the critical temperature of a pure system, 
$\Gamma=n/\pi N_0$, and $n$ is the impurity concentration. 
Based on the impurity effect on $T_c$ we can distinguish two groups of 
scattering potentials determined by the value of the coupling term 
$\left<ef\right>$: the in phase potential with $\left<ef\right>=1$, 
and the out of phase potential defined by $\left<ef\right>=0$. 
Analysis of Eqs. (\ref{e2}) and (\ref{e3}) shows that the impurity pair-breaking 
is minimized for the in phase scattering, that is for $\left<ef\right>=1$ 
or close to $1$, which is achieved for 
$f\left({\bf k}\right)=\sqrt 2\left(k^2_x-k_y^2\right)$ or 
$f\left({\bf k}\right)=sgn\left(k^2_x-k_y^2\right)$. It has been shown that when 
the scattering is close to being in phase it reproduces the 
observed $T_c$ suppression \cite{14,15} and the $H_{c_{2}}$ 
critical field initial slope \cite{11,16} in the cuprates. On the other hand, 
when the scattering is out of phase with the order parameter 
$f\left({\bf k}\right)=2\sqrt 2 k_xk_y$, or 
$f\left({\bf k}\right)=sgn\left(k_xk_y\right)$ it gives 
$\left<ef\right>=0$ and strong suppression of the critical temperature. 
The differing effect of these two groups of impurity potential is also 
seen in the solution to the gap equation. In Fig. 1 we show the gap at a 
temperature of $0.1 T_{c_{0}}$ as a function of the anisotropy degree $\alpha$ 
for in phase (solid line) and out of phase scattering (dashed line) 
in the unitary limit. We note that the anisotropy leading to the maximal $T_c$ 
for the in phase impurity potential, that is $\alpha\sim 0.5$, gives the minimal 
$T_c$ for out of phase scattering.  

\begin{figure}[t]
\parbox{2.7cm}{\vfill $$\frac{\Delta(0.1T_{c_{0}})}{T_{c_{0}}}$$\vfill}
\parbox{5cm}{\includegraphics[height=5cm,width=5cm]{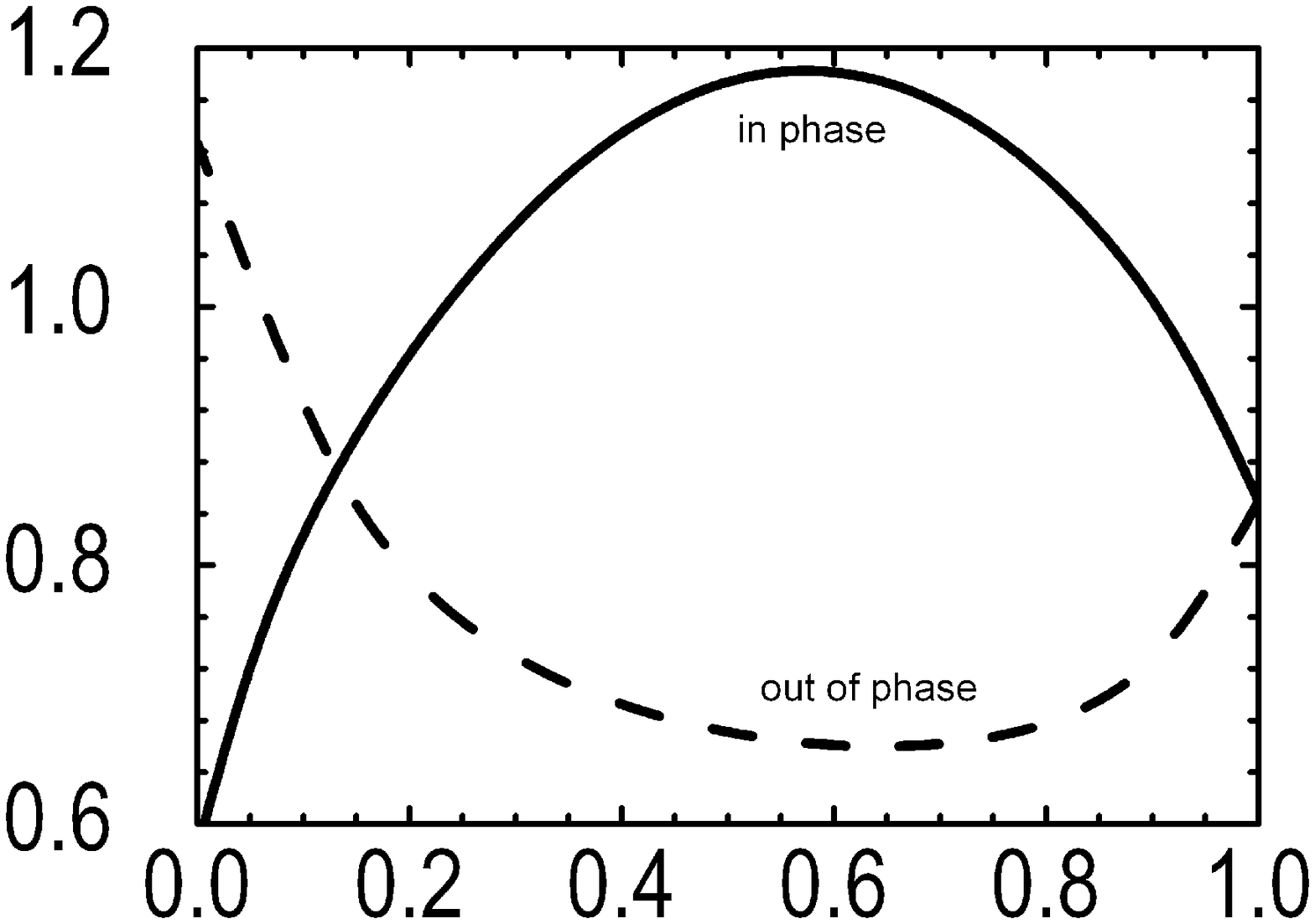}}
\parbox{0.5cm}{\hfill}
\parbox{9.5cm}{\hfill $$\alpha$$\hfill}
\\
\caption{Order parameter vs. partition parameter $\alpha$ for
the in phase scattering (solid line) and  out of phase scattering
(dashed line). Scattering strength $c=0.1/\pi$, impurity concentration
n=0.5 $\Delta_0 N_0$}
\end{figure}

\section{\bf Local density of states}
The position-dependent change of the quasiparticle density of states 
around a single impurity is determined by the real space transform of 
the retarded Green's function $\hat{G}$ and reads 
$\D\delta N\left({\bf r},\omega\right)=
-\frac{\D 1}{\D\pi}Im\left\{\delta G_{11}\left({\bf r},\omega\right)\right\}$. 
The Green's function is obtained from Dyson's equation 
$\hat{G}^{-1}\left(i\omega_n\right)=\hat{G}_0^{-1}\left(i\omega_n\right)
-\hat{\Sigma}$, where $\hat{G}_0$ is the quasiparticle propagator for a pure 
system and the impurity-induced self-energy $\hat{\Sigma}$ is obtained within 
a single impurity approximation, i.e., the t-matrix equation is solved with 
the Green's function of a pure system. 
We have evaluated the LDOS at the impurity site for different partitions of 
the impurity potential starting from isotropic scattering ($\alpha=1$) and ending 
on purely anisotropic scattering ($\alpha=0$). For in phase scattering 
(Fig. 2) we observe a shift of the spectral density from the hole-like 
resonant state for isotropic scattering to the electron-like resonant state 
for scattering close to being purely anisotropic. This effect is similar to the 
changes in the LDOS induced by deviations from particle-hole symmetry, or by 
changes in sign of the scattering potential \cite{17,18}. The influence of the 
out of phase anisotropy of the scattering potential is entirely different. 
It turns out that the increasing degree of anisotropy broadens and finally destroys 
the impurity resonant state (Fig. 3).  

\begin{figure}[t]
\parbox{5cm}{\includegraphics[height=4.3cm,width=5cm]{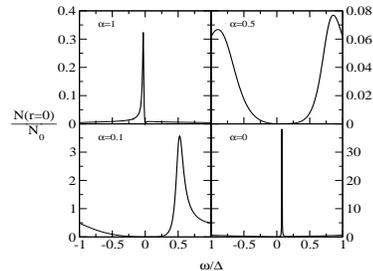}}
\\
\caption{Local density of states at the impurity site (r=0) induced by the
in phase impurity potential,
$f\left({\bf k}\right)=\sqrt 2\left(k^2_x-k_y^2\right)$, for different
partitions of the impurity potential: $\alpha=1$ (isotropic scattering),
$0.5,0.1,0$ (purely anisotropic scattering). Scattering strength $c=0.1$.}
\end{figure}

\begin{figure}[t]
\parbox{5cm}{\includegraphics[height=4.3cm,width=5cm]{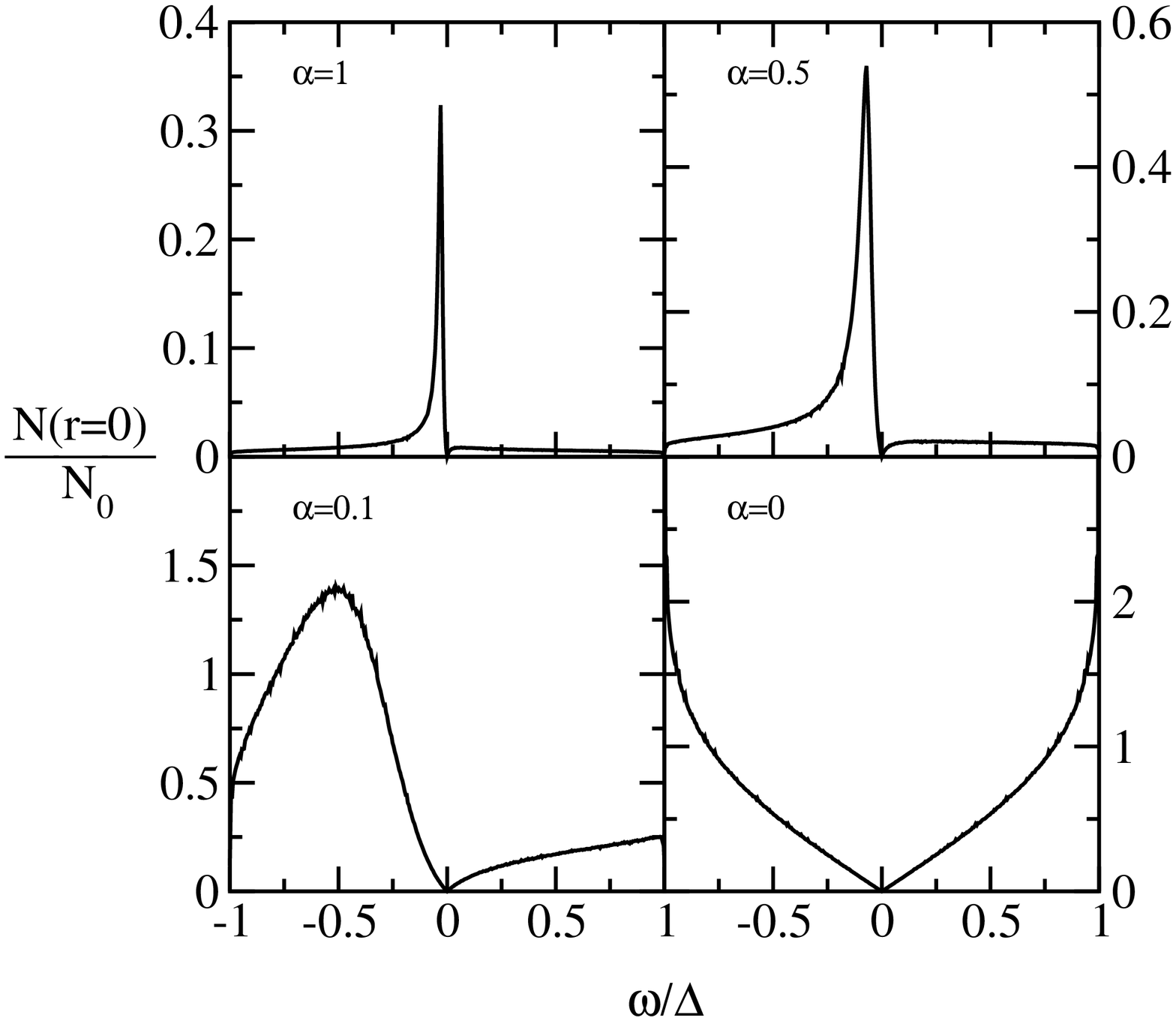}}
\\
\caption{Local density of states at the impurity site (r=0) induced by the
out of phase impurity potential,
$f\left({\bf k}\right)=\sqrt 2k_xk_y$, for different
partitions of the impurity potential: $\alpha=1$ (isotropic scattering),
$0.5,0.1,0$ (purely anisotropic scattering). Scattering strength $c=0.1$.}
\end{figure}

\begin{figure}[h]
\parbox{2.0cm}{\vfill $$\frac{N\left(\omega\right)}{N_0}$$\vfill}
\parbox{5cm}{\includegraphics[height=4.24cm,width=5cm]{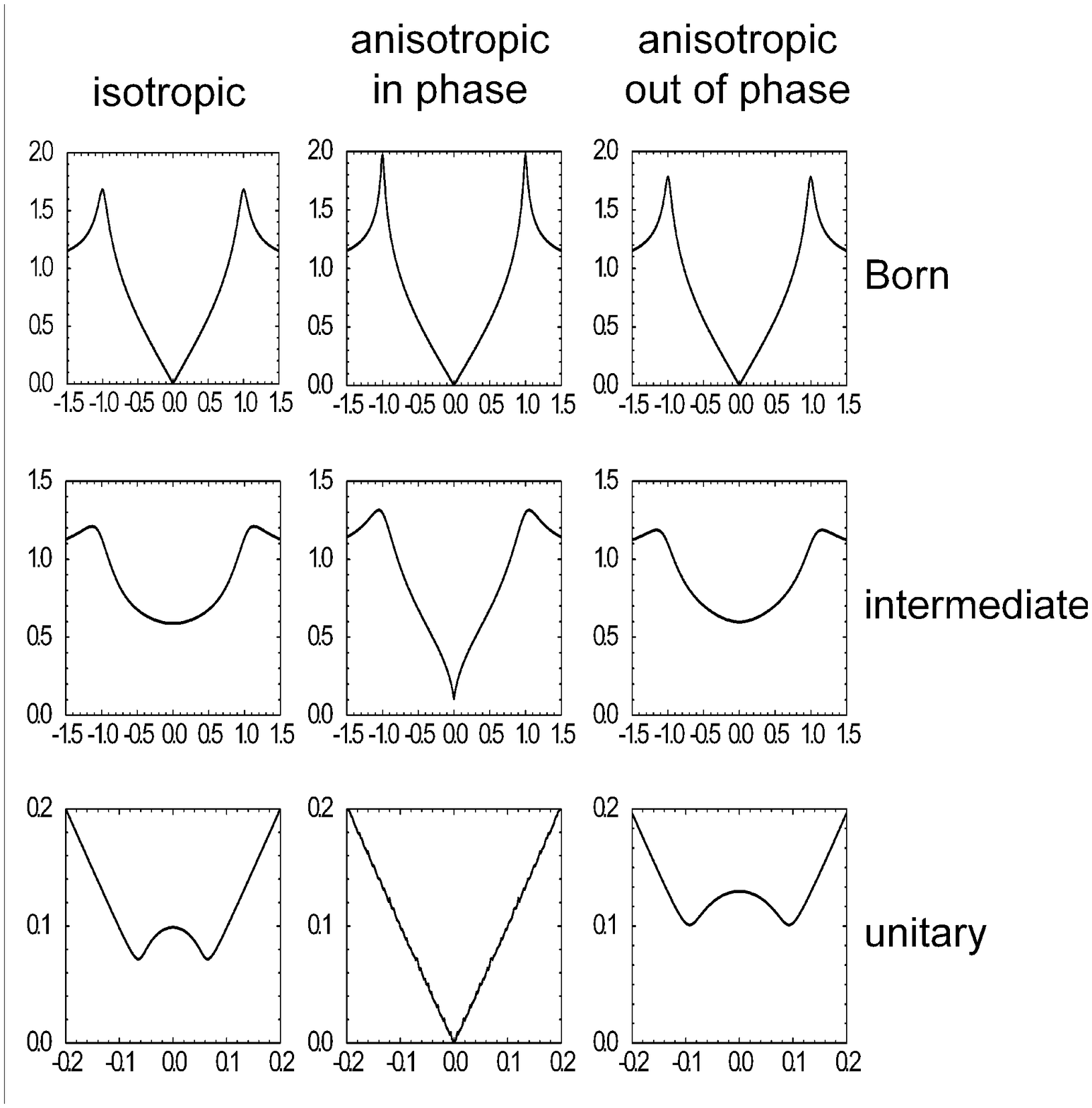}}
\vspace*{1cm}
\parbox{10cm}{\hfill $$\omega/\Delta$$\hfill}
\\
\caption{The density of states for Born scattering
$c=10/\pi$ and impurity concentration n=1 $\Delta_0 N_0$:
a) isotropic potential, b) anisotropic in phase,
c) anisotropic out of phase;
intermediate scattering $c=1/\pi$
and impurity concentration n=1 $\Delta_0 N_0$:
d) isotropic potential, e) anisotropic in phase,
f) anisotropic out of phase;
resonant scattering $c=0.001/\pi$
and impurity concentration n=0.01 $\Delta_0 N_0$:
g) isotropic potential, h) anisotropic in phase,
i) anisotropic out of phase. The partition of the impurity potential
is set to $\alpha=0.8$.}
\end{figure}

\section{\bf Density of states} 
The quasiparticle density of states for a uniform impurity distribution 
is obtained self-consistently within the t-matrix approximation,  
i.e., the matrix $\hat{T}$ is obtained with the use of a dressed 
single-particle propagator $\hat{G}^{-1}=\hat{G}_0^{-1}-\hat{\Sigma}$, 
and the self-energy is determined by $\hat{\Sigma}=n\hat{T}$, 
where $n$ is the impurity concentration \cite{13}. We have evaluated the 
density of states, given by the imaginary part of the retarded Green's function 
$N\left(\omega\right)=-\frac{1}{\pi}Im\sum_{{\bf k}}G_{11}\left({\bf k},\omega\right)$, 
in the limit of Born scattering ($c=10/\pi$), intermediate scattering ($c=1/\pi$) 
and unitary scattering ($c=0.001/\pi$). For each considered scattering 
strength we have compared the effect of the isotropic impurity potential, the anisotropic 
potential in phase with the order parameter and the out of phase anisotropic 
potential (Fig. 4). The degree of anisotropy was set to $\alpha=0.8$. 
Although the discussed potentials lead to similar densities of states in the 
Born limit, observable differences between anisotropic in phase and out of 
phase potentials are seen for intermediate scattering and they become 
significant for the resonant scattering when the in phase potential gives 
$N(\omega=0)=0$, and the out of phase potential enhances the non-zero value 
of $N(\omega=0)$. 

\section{\bf Conclusions}
We have shown that the anisotropy of the impurity potential can 
broaden or destroy the impurity resonant state. The energy level of the 
impurity resonant state relative to Fermi energy can be shifted by the 
anisotropy of the impurity potential, in particular the hole-like resonant 
state is transformed into the electron-like resonant state for highly anisotropic 
in phase scattering. Significant is also effect of anisotropy on the 
quasiparticle density of states which is best seen for unitary impurities. 
The nonzero value of the low-energy density of states for isotropic 
scattering is enhanced by the anisotropic impurity potential which is 
out of phase with the order parameter. On the other hand, the in phase 
impurity potential leads to a vanishing density of states at the Fermi level.  
These results show that the quasiparticle properties of a disordered d-wave 
superconductor depend, apart from the way of modeling the disorder 
\cite{19,3}, on the internal structure of a scattering center. 

\noindent
The work was supported in part by KBN grant No. 5P03B05820.


\begin{thebibliography}{20}
\bibitem{4} S. H. Pan, E. W. Hudson, K. M. Lang, H. Eisaki, S. Uchida,
and J. C. Davis, Nature {\bf 403}, 746 (2000).
\bibitem{5} E. W. Hudson, K. M. Lang, V. Madhavan, S. H. Pan, H. Eisaki,
S. Uchida, and J. C. Davis, Nature {\bf 411}, 920 (2001).
\bibitem{6} J. M. Byers, M. E. Flatt\'e, and D. J. Scalapino, Phys. Rev. Lett.
{\bf 71}, 3363 (1993).
\bibitem{7} M. I. Salkola, A. V. Balatsky, and D. J. Scalapino, Phys. Rev. Lett.{\bf 77}, 1841 (1996).
\bibitem{8} J. Giapintzakis, D. M. Ginsberg, M. A. Kirk, and S. Ockers,
Phys. Rev. B {\bf 50}, 15967 (1994).
\bibitem{9} S. Tolpygo, J. -Y. Lin, M. Gurvitch, S. Y. Hou, and J. M. Phillips,
Phys. Rev. B {\bf 53}, 12454 (1996).
\bibitem{10} S. Tolpygo, J. -Y. Lin, M. Gurvitch, S. Y. Hou, and J. M. Phillips,Phys. Rev. B {\bf 53}, 12462 (1996).
\bibitem{11} J. -Y. Lin, S. J. Chen, S. Y. Chen, C. F. Chang, H. D. Yang,
S. K. Tolpygo, M. Gurvitch, Y. Y. Hsu, and H. C. Ku, Phys. Rev. B {\bf 59},
6047 (1999).
\bibitem{12} G. Hara\'n and A. D. S. Nagi, Phys. Rev. B {\bf 54}, 15463 (1996).
\bibitem{13} G. Hara\'n and A. D. S. Nagi, Phys. Rev. B {\bf 58}, 12441 (1998).
\bibitem{14} G. Hara\'n and A. D. S. Nagi, Phys. Rev. B {\bf 63}, 012503 (2001).
\bibitem{15} G. Hara\'n and A. D. S. Nagi, Acta Phys. Pol. B {\bf 32}, 3459
(2001).
\bibitem{16} H. Won and K. Maki, Physica C {\bf 282-287}, 1837 (1997);
Physica B {\bf 244}, 66 (1998). 
\bibitem{17} A. Polkovnikov, S. Sachdev, and M. Vojta, Phys. Rev. Lett. {\bf 86},
296 (2000).
\bibitem{18} I. Martin, A. V. Balatsky, and J. Zaanen, Phys. Rev. Lett. {\bf 88},
097003 (2002). 
\bibitem{19} W. A. Atkinson, P. J. Hirschfeld, A. H. MacDonald, and K. Ziegler, 
Phys. Rev. Lett. {\bf 85}, 3926 (2000). 
\bibitem{3} A. M. Martin, G. Litak, B. L. Gy\"orffy, J. F. Annett, and
K. I. Wysoki\'nski, Phys. Rev. B {\bf 60}, 7523 (1999). 
\end{thebibliography}
\end{document}